\documentclass[journal]{IEEEtran}
\usepackage{amsmath,amsfonts}
\usepackage{algorithmic}
\usepackage{algorithm}
\usepackage{array}
\usepackage{textcomp}
\usepackage{stfloats}
\usepackage{url}
\usepackage{verbatim}
\usepackage{graphicx}
\usepackage{cite}
\usepackage{amsmath}
\usepackage{amssymb}
\usepackage{epstopdf}
\usepackage{subfigure}
\usepackage[section]{placeins}
\usepackage{graphicx}
\begin{document}

\title{Diffusion Augmented Complex Maximum Total Correntropy Algorithm for Power System Frequency Estimation}

\author{Haiquan Zhao,~\IEEEmembership{Senior Member,~IEEE,}  Yi Peng, Jinsong Chen and Jinhui Hu 
\thanks{This work was partially supported by National Natural Science Foundation of China (grant: 62171388, 61871461, 61571374). Haiquan Zhao, Yi Peng ,  Jinsong Chen and Jinhui Hu are with the School of Electrical Engineering, Southwest Jiaotong University, Chengdu, 610031, China. (e-mail: $hqzhao\_swjtu@126.com$; $pengyi1007@163.com$; $a821719645@163.com$; $jhhu\_swjtu@126.com$)
 
Corresponding author: Haiquan Zhao.
}}



\maketitle

\begin{abstract}
Currently, adaptive filtering algorithms have been widely applied in frequency estimation for power systems. However, research on diffusion tasks remains insufficient. Existing diffusion adaptive frequency estimation algorithms exhibit certain limitations in handling input noise and lack robustness against impulsive noise. Moreover, traditional adaptive filtering algorithms designed based on the strictly-linear (SL) model fail to effectively address frequency estimation challenges in unbalanced three-phase power systems.
To address these issues, this letter proposes an improved diffusion augmented complex maximum total correntropy (DAMTCC) algorithm based on the widely linear (WL) model. The proposed algorithm not only significantly enhances the capability to handle input noise but also demonstrates superior robustness to impulsive noise. Furthermore, it successfully resolves the critical challenge of frequency estimation in unbalanced three-phase power systems, offering an efficient and reliable solution for diffusion power system frequency estimation. Finally, we analyze the stability
 of the algorithm and computer simulations verify the excellent
 performance of the algorithm.
\end{abstract}

\begin{IEEEkeywords}
WL model, frequency estimation, total least squares, augmented, diffusion
\end{IEEEkeywords}

\section{Introduction}
\IEEEPARstart{A}{S} adaptive filter (AF) theory evolves, AF algorithms have been widely applied in system identification, echo cancellation, and channel equalization \cite{LMS}. Nevertheless, traditional AF algorithms, such as the least mean square (LMS) algorithm, are designed in the real domain and are inadequate for the demands of power system frequency estimation. The development of the classical complex LMS (CLMS) algorithm \cite{CLMS} marked a significant advancement, extending traditional real-domain AF methods to the complex domain, thereby enabling accurate frequency estimation for balanced three-phase power systems.

However, general complex algorithms are proposed based on the strictly-linear (SL) model. When the balanced three-phase system deviates from the normal operating state, the SL model-based algorithms will not enable adaptive frequency estimation because the SL model cannot fully characterize the statistics of the complex-valued voltage. To solve this problem, study \cite{ACLMS} employs the widely linear (WL) model to describe complex-valued voltages and proposes the augmented CLMS (ACLMS) algorithm. 

Unfortunately, LMS-based algorithms \cite{ACGMBZ,ACMCC,ACMEE,ACMEEF,zhu2020robust,zhu2021cascaded} have a corresponding performance degradation when applied to the errors-in-variables (EIV) model, where the input signal is also corrupted by noise \cite{EIVmodel}. To overcome this problem, the total least squares (TLS) method was introduced \cite{GDTLS}. In recent years, researchers have proposed the augmented complex TLS (ACTLS) \cite{ACTLS} algorithm to address the frequency estimation problem under the EIV model.  Nonetheless, these algorithms exhibit a lack of robustness, leading to significant performance degradation under impulsive noise interference in the output. To overcome this limitation, the augmented complex-valued minimum total error entropy with
fiducial points (ACMTEEF) algorithm for frequency estimation has been introduced \cite{ACMTEEF}.

Furthermore, the existing algorithms are designed for single-node frameworks, and research on distributed approaches remains scarce \cite{DACLMS,DACISR}. As far as we know, no algorithm has yet been developed to address all these challenges concurrently, which  motivates this work. 

Considering the advantages of diffusion networks, including faster convergence and lower steady-state error, in this letter, we introduce the maximum correntropy criterion (MCC) criterion into the TLS algorithm,  resulting in the diffusion augmented complex maximum total correntropy (DAMTCC) algorithm within the  WL model. The DAMTCC algorithm effectively inherits the robustness of the MCC criterion against impulsive noise and the TLS algorithm's capability to handle input noise. Besides, this algorithm provides accurate frequency estimation performance  when the system deviates from its normal operating state. Moreover, this letter also carefully analyzes the convergence of the algorithm and derives the step size range that ensures the stable operation of the algorithm. Finally, the excellent performance of the algorithm is verified by computer simulation.

\section{Problem Description}
\subsection{Complex-valued voltage signal for frequency estimation }
The three-phase voltage signals can be represented as
\begin{equation}\label{eq1}
\begin{aligned}
{v_a}(\tau) &= {V_a}(\tau)\cos \left( {w\tau \Delta T + \vartheta    } \right) \\
{v_b}(\tau) & = {V_b}(\tau)\cos \left( {w\tau \Delta T + \vartheta +  \vartheta _b -2/3\pi  } \right) \\
{v_c}(\tau) & = {V_c}(\tau)\cos \left( {w\tau \Delta T + \vartheta +  \vartheta _c +2/3\pi  } \right) \\
\end{aligned}
\end{equation}
here, ${V_a}(\tau)$, ${V_b}(\tau)$, ${V_c}(\tau)$ are voltage amplitudes of the corresponding phases,  at time instant $\tau$, respectively. $w=2{\pi}f $ is the angular frequency and $f$ is system frequency. $\Delta T$ denotes the sampling interval, $\vartheta $ is the phase of the initial angle. $\vartheta _b$ and $\vartheta_c$ are initial phase difference from phase $a$, respectively. Using the Clark's transformation \cite{17}, the three-phase signals ${v_a}(\tau)$, ${v_b}(\tau)$, ${v_c}(\tau)$ can be mapped into the $\alpha\beta0$ coordinate system as
\begin{equation}\label{eq2}
\left[ {\begin{array}{*{20}{c}}
{{v_0}}\\
{{v_\alpha }}\\
{{v_\beta }}
\end{array}} \right] = \sqrt {\frac{2}{3}} \left[ {\begin{array}{*{20}{c}}
{\frac{{\sqrt 2 }}{2}}&{\frac{{\sqrt 2 }}{2}}&{\frac{{\sqrt 2 }}{2}}\\
1&{ - \frac{1}{2}}&{ - \frac{1}{2}}\\
0&{\frac{{\sqrt 3 }}{2}}&{ - \frac{{\sqrt 3 }}{2}}
\end{array}} \right]\left[ {\begin{array}{*{20}{c}}
{{v_a}(\tau)}\\
{{v_b}(\tau)}\\
{{v_c}(\tau)}
\end{array}} \right]
\end{equation}
The factor $\sqrt{2/3} $  \ is produced to keep the power constant under this variation. Typically, no analysis is required for the 0 sequence, and only the $\alpha$ and $\beta$ components are utilized in the model \cite{18}. Therefore, it is easy to construct a complex-valued voltage signal for frequency estimation as
\begin{equation}\label{eq3}
v(\tau) = {v_\alpha }(\tau) + j{v_\beta }(\tau)
\end{equation}

Combining (\ref{eq2}) and (\ref{eq3}), then (\ref{eq3}) can be further expressed as
\begin{equation}\label{eq4}
v(\tau) = A(\tau)\exp \left( {j\left( {w\tau\Delta T + \vartheta } \right)} \right) + B(\tau)\exp \left( { - j\left( {w\tau\Delta T + \vartheta } \right)} \right)
\end{equation}
where, $A(\tau) = \frac{{\sqrt 6 }}{6}\left( {{V_a}(\tau) + {V_b}(\tau){e^{j\Delta {\vartheta _b}}} + {V_c}(\tau){e^{j\Delta {\vartheta _c}}}} \right)$ and $B(\tau) = \frac{{\sqrt 6 }}{6}\left( {{V_a}(\tau) + {V_b}(\tau){e^{ - j\left( {\Delta {\vartheta _b} + \frac{{2\pi }}{3}} \right)}} + {V_c}(\tau){e^{ - j\left( {\Delta {\vartheta _c} - \frac{{2\pi }}{3}} \right)}}} \right)$

In case of the balanced three-phase power system, we have $\Delta {\vartheta _b} =\Delta {\vartheta _c} = 0$ and ${V_a}(\tau) = {V_b}(\tau) = {V_c}(\tau)$ , i.e.
\begin{equation}\label{eq7}
A(\tau) = \sqrt {\frac{3}{2}} {V_a}(\tau), B(\tau) = 0
\end{equation}

Substituting (\ref{eq7}) into (\ref{eq4}) gives
\begin{equation}\label{eq9}
v(\tau) = \sqrt {\frac{3}{2}} {V_a}(\tau)\exp \left [{j\left( {w\tau\Delta T + \vartheta } \right)} \right]
\end{equation}
then,
\begin{equation}
\begin{split}
v(\tau+1) &= \sqrt {\frac{3}{2}} {V_a}(\tau+1)\exp \left [{j\left( {w(\tau+1)\Delta T + \vartheta } 
\right)} \right]\\
&=\sqrt {\frac{3}{2}} {V_a}\exp \left [{j\left( {w(\tau)\Delta T + \vartheta } \right)} \right]\exp \left [{j\left( {w(\Delta T  } 
\right)} \right]\\
&=v(\tau)\exp \left [{j\left( {w(\Delta T  } \right)} \right]\label{eq10}
\end{split}
\end{equation}

From (\ref{eq10}), we can know that $v(\tau)$ is second-order circular, since its probability density function is rotation invariant \cite{ACLMS}. Thus, in a balanced three-phase power system, frequency estimation can be realized by the SL model.

\subsection{EIV Model}
Considering a linear system:
\begin{equation}
    {d}(\tau)=\textbf{{x}}^{H}\textbf{{w}}^o(\tau) \label{eq12}
\end{equation}
where $\textbf{{w}}^o(\tau)\in {\mathbb{C}}^{L\times 1} $ is the unknown weight vector,  $ \textbf{\emph{x}}(\tau)$ is the complex-values input vector at time instant $\tau$ and ${d}(\tau)$ is corresponding output signal.
\begin{equation}
\tilde{\textbf{\emph{x}}}(\tau)={\textbf{\emph{x}}} (\tau)+{\textbf{\emph{m}}} (\tau) \label{eq13}
\end{equation}
\begin{equation}
\tilde{d}(\tau)={d} (\tau)+{n} (\tau) \label{eq14}
\end{equation}
where ${\textbf{\emph{m}}} (\tau) \in {\mathbb{C}^{L\times 1}}$ and $ {n} (\tau)\in {\mathbb{C}} $ are the input and output Gaussian white noise signals, and the variances of their real and imaginary parts are $ {\sigma}_{i}^{2}\textbf{I}/2$ and ${\sigma}_{o}^{2}/2$, respectively \cite{ACMTEEF}.

\section{The Proposed Algorithms for Frequency Estimation}
In practice, the three-power system may transform an unbalanced three-power system. In this case, $B(\tau) \ne  0$, so the estimated value of voltage and estimation error are derived for frequency estimation by WL model as
\begin{equation}
    \hat{v}_{u,l}(\tau+1)=v_{u,l}(\tau)\emph{h}_{u,l}^*(\tau)+v_{u,l}^{*}(\tau)\emph{g}_{u,l}^*(\tau) \label{eq24}
\end{equation}
\begin{equation}\label{eq25}
    e_{u,l}(\tau)=v_{u,l}(\tau+1)-v_{u,l}(\tau)\emph{h}_{u,l}^*(\tau)-v_{u,l}^{*}(\tau)\emph{g}^*_{u,l}(\tau)
\end{equation}
where $v_{u,l}(\tau)$ is the unbalanced complex-value voltage signal at node $l$ at time instant $\tau$, $\hat{v}_{u,l}(\tau+1)$ is the estimated voltage, and $h_{u,l}(\tau)$ and $\emph{g}_{u,l}(\tau)$ are the standard and conjugate weight coefficients, respectively.

To simplify the derivation process, (\ref{eq25}) is further expressed as
\begin{equation}\label{23}
    e_{u,l}(\tau)=v_{u,l}(\tau+1)-\textbf{w}_{l}(\tau)^H\textbf{x}_{l}(\tau)
\end{equation}
where $\textbf{w}_{l}(\tau)\triangleq \left[\emph{h}_{u,l}(\tau),\emph{g}_{u,l}(\tau)\right]^T$ and $\textbf{x}_{l}(\tau)\triangleq\left[v_{u,l}(\tau),v_{u,l}^*(\tau)\right]^T$

The cost function of DAMTCC algorithm is denoted as
\begin{equation}\label{eq26}
J_{{DAMTCC,l}}=E\left[ {\exp \left( { - \frac{{{{\left| {e_{u,l}\left( \tau\right)} \right|}^2}}}{{2{\sigma ^2}{\left | h_{u,l}\right |^2+\left | \emph{g}_{u,l}\right |^2+\gamma }}}} \right)} \right]
\end{equation}

Then, the standard and conjugate weight coefficients update formula can be denoted separately as
\begin{equation}\label{eq27}
\begin{aligned}
    {\varPsi }_{u,l}(\tau+1)&=\emph{h}_{u,l}(\tau)+\mu_{u,l}\nabla_{\emph{h}_{u,l}^{*}}J_{DAMTCC,l}\\
    \varUpsilon_{u,l}(\tau+1)&=\emph{g}_{u,l}(\tau)+\mu_{u,l}\nabla_{\emph{g}_{u,l}^{*}}J_{DAMTCC,l}
\end{aligned}
\end{equation}
where  $\mu_{u,l}$ is the step-size at node $l$,  $\nabla _{h_{u,l}^*}J_{DAMTCC,l}$ and $\nabla _{\emph{g}_{u,l}^*}J_{DAMTCC,l}$ are the two instantaneous complex gradients of the cost function, which can be written as
\begin{equation}\label{eq28}
    \begin{split}
    &\nabla _{h_{u,l}^*}J_{DAMTCC,l}=\exp(-\frac{\left | e_{u,l}(\tau)\right |^2}{2\sigma ^2(\left | h_{u,l}\right |^2+\left | \emph{g}_{u,l}\right |^2+\gamma )})\\
    &\times\frac{e_{u,l}^*(\tau )v_{u,l}(\tau )(\left | h_{u,l}\right |^2+\left | \emph{g}_{u,l}\right |^2+\gamma )+\left | e_{u,l}\right |^{2}h_{u,l}(\tau)}{2\sigma ^2(\left | h_{u,l}\right |^2+\left | \emph{g}_{u,l}\right |^2+\gamma )^{2}}
    \end{split}
\end{equation}
\begin{equation}\label{eq29}
    \begin{split}
    &\nabla _{g_{u,l}^*}J_{DAMTCC,l}=\exp(-\frac{\left | e_{u,l}(\tau)\right |^2}{2\sigma ^2(\left | h_{u,l}\right |^2+\left | \emph{g}_{u,l}\right |^2+\gamma )})\\
    &\times\frac{e_{u,l}^*(\tau )v_{u,l}^*(\tau )(\left | h_{u,l}\right |^2+\left | \emph{g}_{u,l}\right |^2+\gamma )+\left | e_{u,l}\right |^{2}\emph{g}_{u,l}(\tau)}{2\sigma ^2(\left | h_{u,l}\right |^2+\left | \emph{g}_{u,l}\right |^2+\gamma )^{2}}
    \end{split}
\end{equation}
here, we utilize the instantaneous value of the gradient instead of the expectation value, which is common in the AF domain.

Then, the updating formulas for the global standard weight coefficient and the conjugate weight coefficient are given separately as
\begin{equation}\label{eq32}
\begin{aligned}
    {\emph{h}}_{u,l}(\tau+1)&=\displaystyle\sum_{i\in N_{l}}c_{i,l}\varPsi _{u,i}(\tau+1)\\
    {\emph{g}}_{u,l}(\tau+1)&=\displaystyle\sum_{i\in N_{l}}c_{i,l}\varUpsilon_{u,i}(\tau+1)
\end{aligned}
\end{equation}

Given that the system frequency is significantly lower than the sampling frequency, the imaginary part of $e^{j\hat{w}\varDelta T}$ is positive definite \cite{DACISR}. Therefore, the frequency estimate for the unbalanced three-power system can be derived as 
\begin{equation}\label{eq33}
    \hat{f}_{u,l}=\arcsin{(\mathfrak{I}( {\emph{h}}_{u,l}(\tau)+a(\tau)\emph{g}_{u,l}(\tau)))}/2\pi\varDelta T
\end{equation}
where
\begin{equation}\label{eq34}
    a(\tau)=\left ({-j\mathfrak{I}(h_{u,l}(\tau))+j\sqrt{\mathfrak{I}^2(h_{u,l}(\tau))-\left | \emph{g}_{u,l}(\tau)\right |^2}}\right )/{g_{u,l}(\tau )}
\end{equation}
and $\mathfrak{I}(\cdot)$ denotes the imaginary part of a complex number.

Thus, the implementation procedure of DAMTCC algorithm can be see in Algorithm 1.
\begin{algorithm}[!h]
	\caption{DAMTCC}
    \label{alg:Sum}
    \begin{algorithmic}[1]
    
    	\STATE Parameters $\sigma$, $\mu_{u,l}$
      	\FOR{$\tau=1,2,3 ...$}
            \FOR{each node $l$}
                \STATE Updated
        	\STATE $e_{u,l}(\tau)=v_{u,l}(\tau+1)-v_{u,l}(\tau)\emph{h}_{u,l}^*(\tau)-v_{u,l}^{*}(\tau)\emph{g}^*_{u,l}(\tau)$
                \STATE ${\varPsi }_{u,l}(\tau+1)=\emph{h}_{u,l}(\tau)+\mu_{u,l}\nabla_{\emph{h}_{u,l}^{*}}J_{DAMTCC,l}$
                \STATE
                $\varUpsilon_{u,l}(\tau+1)=\emph{g}_{u,l}(\tau)+\mu_{u,l}\nabla_{\emph{g}_{u,l}^{*}}J_{DAMTCC,l}$
                \STATE Combination
                \STATE $ {\emph{h}}_{u,l}(\tau+1)=\displaystyle\sum_{i\in N_{l}}c_{i,l}\varPsi _{u,i}(\tau+1)$
                \STATE ${\emph{g}}_{u,l}(\tau+1)=\displaystyle\sum_{i\in N_{l}}c_{i,l}\varUpsilon_{u,i}(\tau+1)$
                \STATE Frequency estimation
                \STATE $\hat{f}_{u,l}=\arcsin{(\mathfrak{I}( {\emph{h}}_{u,l}(\tau)+a(\tau)\emph{g}_{u,l}(\tau)))}/2\pi\varDelta T$
                \STATE $    a(\tau)=\frac{\left ({-j\mathfrak{I}(h_{u,l}(\tau))+j\sqrt{\mathfrak{I}^2(h_{u,l}(\tau))-\left | \emph{g}_{u,l}(\tau)\right |^2}}\right )}{{g_{u,l}(\tau )}}$
      	\ENDFOR
            \ENDFOR
    
    \end{algorithmic}
\end{algorithm}
\vspace{-5mm}
\section{Performance Analysis}
In this subsection, we will analyse the convergence performance of the DAMTCC algorithm based on some common assumptions \cite{ACMTEEF,16} as follow:

A1: At each node $l$, input signal $\textbf{x}_{l}(\tau)$, input noise $\textbf{m}_{l}(\tau)$, weight coefficient $\textbf{w}_{l}(\tau)$ and output noise ${n}_{l}(\tau)$ are independent of each other.

A2: The matrix $\textbf{R}_{l}\triangleq E\left[\textbf{x}_{l}(\tau)\textbf{x}_{l}(\tau)^H\right]$ is positive definite full rank.

To facilitate the subsequent derivation, from (\ref{23}), (\ref{eq28}), and (\ref{eq29}), the following expression is given as
\begin{equation}\label{eq35}
    \varPsi ^{c}_{l}(\tau+1)=\textbf{w}_{l}(\tau)+\eta_{l} \textbf{G}_{l}(\tau)
\end{equation}
\begin{equation}\label{32}
    \textbf{w}_{l}(\tau+1)=\displaystyle\sum_{i\in N_{l}}c_{i,l} \varPsi ^{c}_{i}(\tau+1)
\end{equation}
here,  $\textbf{G}_{l}(\tau)=\frac{1}{2\sigma ^2}\exp(-\frac{\left | \bar e_{l}(\tau)\right |^2}{2\sigma ^2{\left \| \bar{\textbf{w}}(\tau)\right \|^2}})(\frac{\bar e^*_{l} \bar{\textbf{x}}(\tau)}{\left \|\bar{\textbf{w}}_{l}(\tau) \right \|^2}+\frac{\left | \bar e_{l}(\tau)\right |^2\textbf{w}_{l}(\tau)}{\left \| \bar{\textbf{w}}_{l}(\tau)\right \|^4} )$, $\varPsi ^{c}_{l}(\tau+1)=[{\varPsi }_{l}(\tau+1),\varUpsilon_{l}(\tau+1)]^T$, $\bar e_{l}(\tau)=e_{l}(\tau)+n_{l}(\tau)$, $\bar{\textbf{x}}_{l}(\tau)={\textbf{x}}_{l}(\tau)+\textbf{m}_{l}(\tau)$, and $\bar{\textbf{w}}_{l}(\tau)=[ {\emph{h}}_{l}(\tau), {\emph{g}}_{l}(\tau),\sqrt{\gamma}]^H$.

As shown in (\ref{eq28}), (\ref{eq29}), and (\ref{eq35}) the update formula of the algorithm usually approximates the expectation with the instantaneous value, but this approximation produces a gradient error, which is denoted as
\begin{equation}\label{eq48}
    \mathcal{G}_{l}(\tau)=\textbf{G}_{l}(\tau)-\mathrm{E}(\textbf{G}_{l}(\tau))
\end{equation}

Substituting (\ref{eq48}) into (\ref{eq35}) gives
\begin{equation}\label{33}
    \varPsi ^{c}_{l}(\tau+1)=\textbf{w}_{l}(\tau)+\eta_{l}\left[E(\textbf{G}_{l}(\tau))+\mathcal{G}_{l}(\textbf{w}(\tau))\right]
\end{equation}

To continue calculating $E(\textbf{G}_{l}(\tau))$, it is first necessary to calculate the Hessian matrix \cite{MTCC} of $J_{{DAMTCC,l}}$ at  $\textbf{w}^o$ as 
\begin{equation}
    \textbf{H}_{DAMTCC,l}(\textbf{w}^o)=\begin{bmatrix}
 \textbf{H}_{A,l}(\textbf{w}^o)& \textbf{H}_{B,l}(\textbf{w}^o)\\ 
\textbf{H}_{B,l}^*(\textbf{w}^o) & \textbf{H}_{A,l}^*(\textbf{w}^o)
\end{bmatrix}
\end{equation}
where $\textbf{H}_{A,l}(\textbf{w}^o)=\frac{\partial^2J_{{DAMTCC,l}}}{\partial \textbf{w}^*\partial \textbf{w}^T}(\textbf{w}^o)=-\frac{1}{2\sigma ^2\left \| \bar{\textbf{w}}^o_{l}(\tau)\right \|^2}\kappa^2 \textbf{R}_{l}$, $\kappa=\frac{\sigma^2}{\sigma^2+\sigma_{i,l}^2/2}$, and $\textbf{H}_{B,l}({\textbf{w}^{o}})=\frac{\partial^2J_{{DAMTCC,l}}}{\partial \textbf{w}^*\partial \textbf{w}^H}(\textbf{w}^o)=0$.

Taking the Taylor formula at $\textbf{w}^{o}$ and defining $    \widetilde{\textbf{w}}_{l}(\tau)\mathop {\rm{ = }}\limits^{{\rm{def}}}\textbf{w}^{o}(\tau)-\textbf{w}_{l}(\tau)$ , $E(\textbf{G}_{l}(\tau))$ can be written as 
\begin{equation}\label{51}
\begin{split}
    E(\textbf{G}_{l}(\tau))& \approx E(\textbf{G}_{l}({\textbf{w}^{o}}))-\textbf{H}_{A,l}({\textbf{w}^{o}})\widetilde{\textbf{w}}_{l}(\tau))\\
    & \approx -\textbf{H}_{A,l}({\textbf{w}^{o}})\widetilde{\textbf{w}}_{l}(\tau)
\end{split}
\end{equation} 
where 
$E(\textbf{G}_{l}({\textbf{w}^{o}}))=E\left[\exp(-\frac{\left |\bar{e}_{l,o} \right |^2}{2\sigma^2{\left \| \bar{\textbf{w}}^o(\tau)\right \|^2}})(\frac{\left |\bar{e}_{l,o} \right |\textbf{m}_{l}(\tau)}{{\left \| \bar{\textbf{w}}^{o}(\tau)\right \|^2}})\right]-E\left[\exp(-\frac{\left |\bar{e}_{l,o} \right |^2}{2\sigma^2{\left \| \bar{\textbf{w}}^o(\tau)\right \|^2}})(\frac{\left |\bar{e}_{l,o} \right |^2\textbf{w}^{o}}{{\left \| \bar{\textbf{w}}^o(\tau)\right \|^4}})\right]=2\kappa \sigma_{i,l}^2\left \|\bar{\textbf{w}}^o(\tau) \right \|^2-2\kappa \sigma_{i,l}^2\left \|\bar{\textbf{w}}^o(\tau) \right \|^2=0$, and $\bar{e}_{l,o}=n_{l}(\tau)-\textbf{w}^{oH}\textbf{m}_{l}(\tau)$.

Substituting (\ref{51}) into (\ref{33}) yields
\begin{equation}
        \varPsi ^{c}_{l}(\tau+1)=\textbf{w}_{l}(\tau)-\eta_{l}\left[\textbf{H}_{A,l}({\textbf{w}^{o}})\widetilde{\textbf{w}}_{l}(\tau)-\mathcal{G}_{l}(\textbf{w}^o)\right]
\end{equation}

Then, we can get the weight vector at  the proximity of $\textbf{w}^{o}$
\begin{equation}
    \textbf{w}_{l}(\tau+1)=\displaystyle\sum_{i\in N_{l}}c_{i,l}\left[ \textbf{w}_{i}(\tau)-\eta_{i}\left[\textbf{H}_{A,i}({\textbf{w}^{o}})\widetilde{\textbf{w}}_{i}(\tau)-\mathcal{G}_{l}(\textbf{w}^o)\right]\right]
\end{equation}

In the case of a network with N nodes, the order of the filter at each node is 2 because the input and weight are augmented vectors. For the purposes of the following analysis, the following definitions are provided, $\mathcal{U}\triangleq diag[\eta_{1}\textbf{I}_{2\times 2},\dots,\eta_{N}\textbf{I}_{2\times 2}]$, $\mathcal{W}^O \triangleq [\textbf{w}^o,\dots,\textbf{w}^o]$, $\mathcal{W}(\tau) \triangleq [\textbf{w}_{1}(\tau),\dots,\textbf{w}_{N}(\tau)]$, $\widetilde{\mathcal{W}}(\tau) \triangleq [\widetilde{\textbf{w}}_{1}(\tau),\dots,\widetilde{\textbf{w}}_{N}(\tau)]$, $\mathfrak{G} \triangleq [\mathcal{G}_{1}(\textbf{w}^o),\dots,\mathcal{G}_{N}(\textbf{w}^o)]$, $\mathcal{C} \triangleq C^T\otimes \textbf{I}_{N}$, and $\mathcal{H} \triangleq diag[\textbf{H}_{A,1}({\textbf{w}^{o}}),\dots,\textbf{H}_{A,N}({\textbf{w}^{o}})]$. The matrix $C$ contains all the combination coefficients ${c_{i,l}}$, thus each column of $C$ sums up to one \cite{DCTLMM}.

With these definitions, (\ref{32}) in the vicinity of the ${\textbf{w}^o}$  can be rewritten as
\begin{equation}\label{39}
 \mathcal{W}(\tau+1)= \mathcal{C}\left[\mathcal{W}(\tau)-\mathcal{U}\mathcal{H}\widetilde{\mathcal{W}}(\tau)+\mathcal{U}\mathfrak{G}\right]  
\end{equation}

Subtracting $\mathcal{W}^O$ from both sides of (\ref{39}), the global weight error vector can be expressed as
\begin{equation}\label{40}
    \widetilde{ \mathcal{W}}(\tau+1)= \mathcal{C}\left[\textbf{I}_{2 N}+\mathcal{U}\mathcal{H}\right]\widetilde{\mathcal{W}}(\tau)-\mathcal{C}\mathcal{U}\mathfrak{G}
\end{equation}

Performing the expectation operation on (\ref{40}), we have
\begin{equation}
\begin{split}
 E\left[\widetilde{ \mathcal{W}}(\tau+1)\right]&= \mathcal{C}\left[\textbf{I}_{2 N}+\mathcal{U}\mathcal{H}\right]E\left[\widetilde{ \mathcal{W}}(\tau)\right]-\mathcal{C}\mathcal{U}E\left[\mathfrak{G}\right]\\
 &=\mathcal{C}\left[\textbf{I}_{2 N}+\mathcal{U}\mathcal{H}\right]E\left[\widetilde{ \mathcal{W}}(\tau)\right]
\end{split}  
\end{equation}

To ensure the stability of the DAMTCC algorithm, it is required to satisfy $\rho  ( \mathcal{C}\left[\textbf{I}_{2 \times N}+\mathcal{U}\mathcal{H}\right] )<1$, where $\rho  ( \cdot)$ is the spectral radius, and $\rho  ( \textbf{C})=1$. Therefore, $\eta_{l}$ should satisfy the condition $\left |1+\eta_l\lambda_{max}[\textbf{H}_{A,l}({\textbf{w}^{o}})] \right |$ have to be less than 1, i.e. $\left |1+\frac{\eta_l\kappa^2 \lambda_{max}(\textbf{R}_{l})}{2\sigma ^2\left \| \bar{\textbf{w}}^o_{l}(\tau)\right \|^2} \right |<1$, and  $\mu_{l}=\frac{\eta_{l}}{2\sigma ^2}$.
Finally, to ensure the stability of the algorithm, the step size can be given as
\begin{equation}
    0< \mu_{l}< -\frac{2}{2\sigma^2\lambda_{min}[\textbf{H}_{A,l}({\textbf{w}^{o}})]}=\frac{{2\left \| \bar{\textbf{w}}^o_{l}(\tau)\right \|^2}}{\kappa^2\lambda_{min}({\textbf{R}_{l}})}
\end{equation}

\section{Simulation}

In this section, the simulation of unbalanced power systems is achieved by D-type voltage sags \cite{19}. The simulation for the performance validation of the proposed DAMTCC algorithms are based on two different topologies with the node $N=8$, which is shown in Fig. 1. As for the noise environment, both input and output noise are Gaussian noise with different signal-to-noise ratios (SNRs) at different nodes, as illustrated in Fig. 2. Moreover, the impulsive noise is generated by a Bernoulli Gaussian process with probability $p=0.005$ and variance $\sigma_{imp}^2=10$. The frequency of the three-phase balanced power system and the sampling frequency of the voltage signals are $50$Hz and $2.5k$Hz, separately. In addition, Metropolis rules \cite{16} are used to generate the combination coefficients $c_{i,l}$. 

\begin{figure}[htbp]
    \centering
    \subfigure[]{
    	\begin{minipage}[b]{0.47\linewidth}
        \centering
        \includegraphics[scale=0.3]{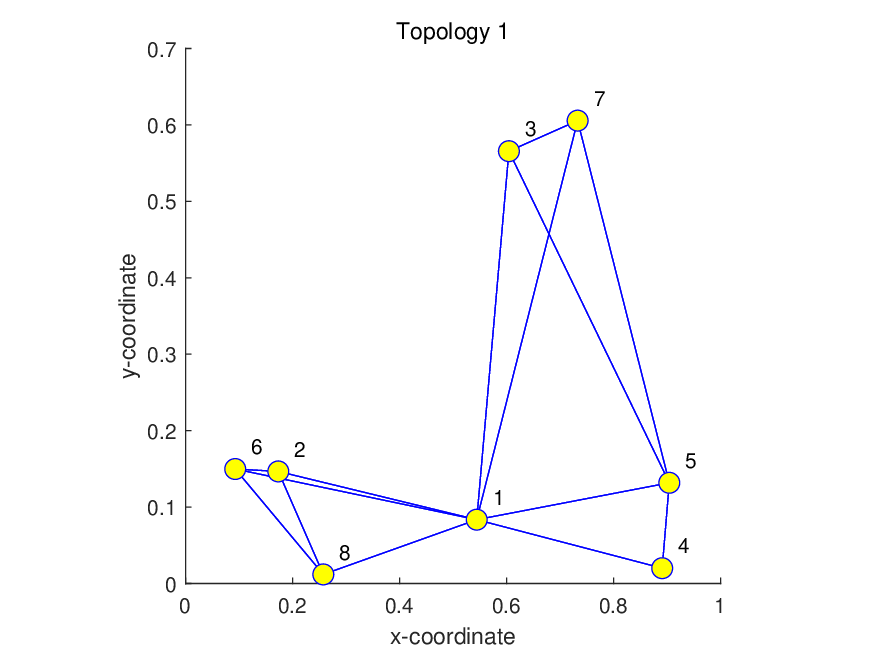} 
    \end{minipage}
    }
        \subfigure[]{
    	\begin{minipage}[b]{0.45\linewidth}
        \centering
        \includegraphics[scale=0.3]{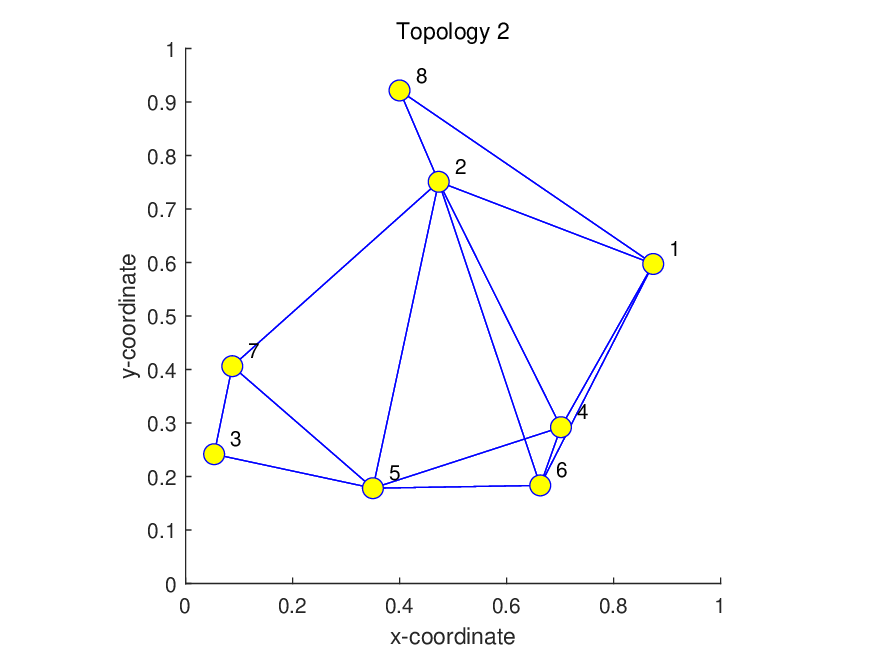}
    \end{minipage}
    }
    \caption{Network Topolog}
    \vspace{-3mm}
\end{figure}

\begin{figure}[htbp]
    \centering
    \subfigure[]{
    	\begin{minipage}[b]{0.46\linewidth}
        \centering
        \includegraphics[scale=0.27]{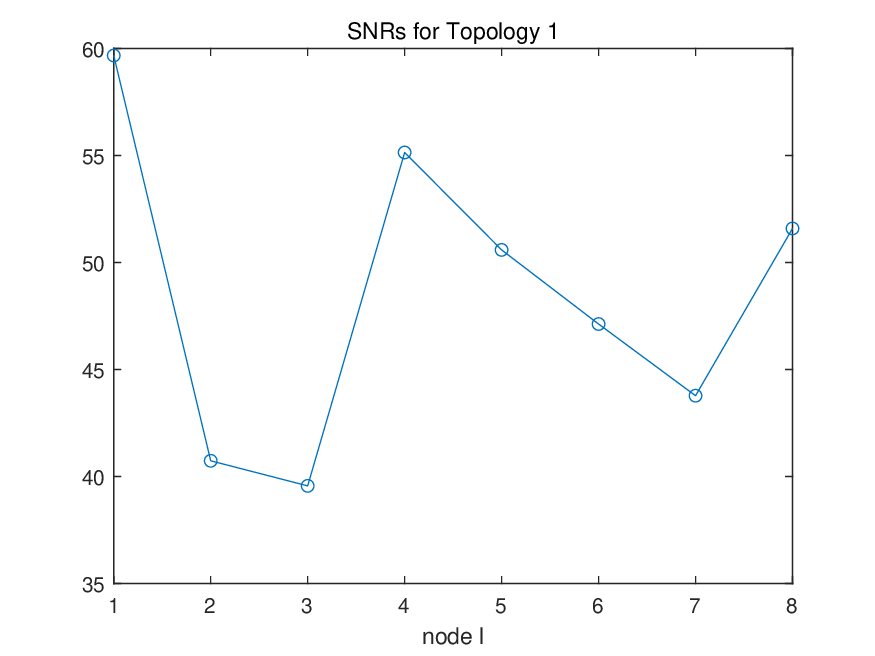} 
    \end{minipage}
    }
        \subfigure[]{
    	\begin{minipage}[b]{0.47\linewidth}
        \centering
        \includegraphics[scale=0.27]{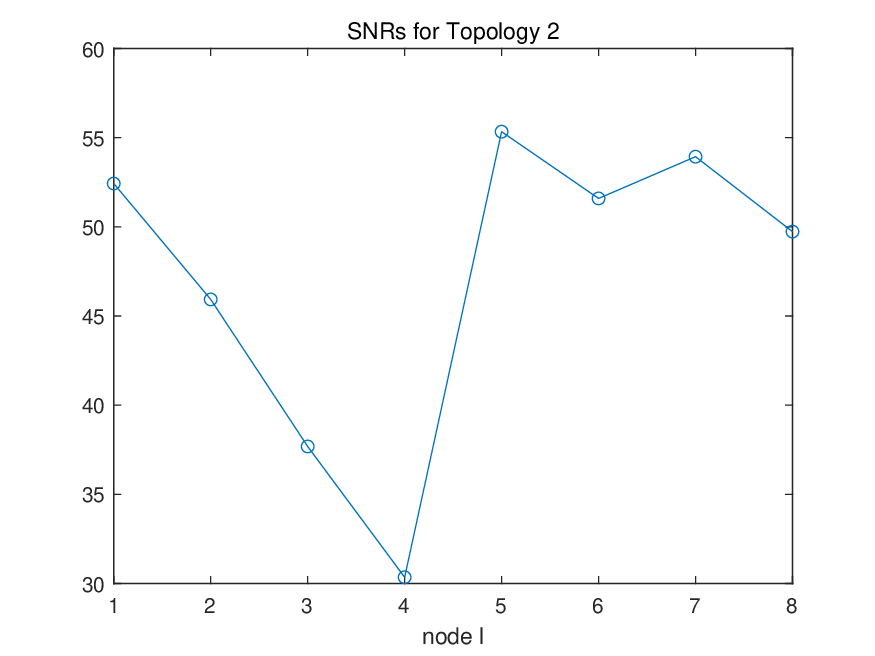}
    \end{minipage}
    }
    \caption{SNRs for different Topologys}
    \vspace{-5mm}
\end{figure}

\begin{figure}[htbp]
    \centering
    \subfigure[]{
    	\begin{minipage}[b]{.46\linewidth}
        \centering 
        
        \includegraphics[scale=0.27]{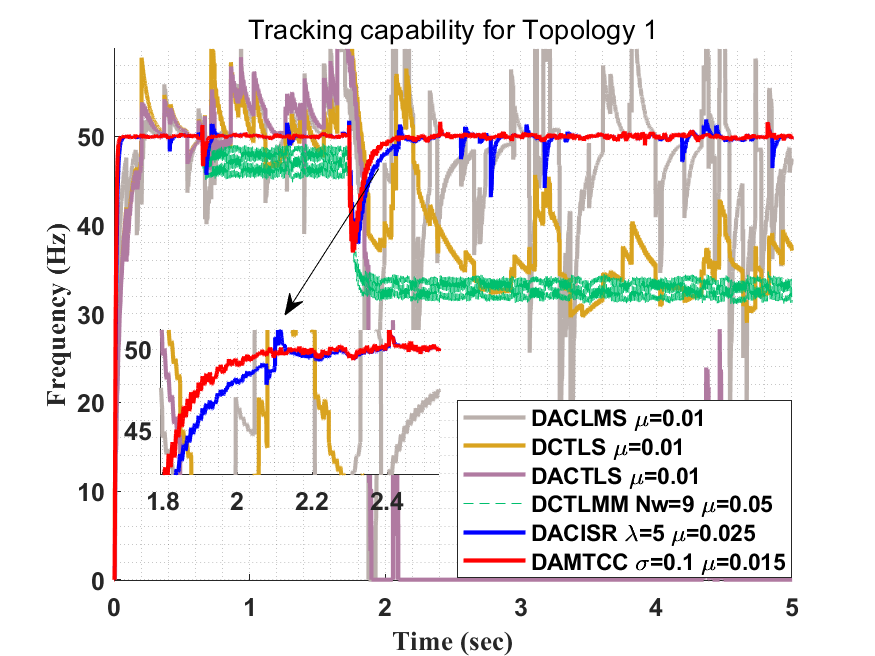} 
    \end{minipage}
    }
        \subfigure[]{
    	\begin{minipage}[b]{.47\linewidth}
        \centering
        \includegraphics[scale=0.27]{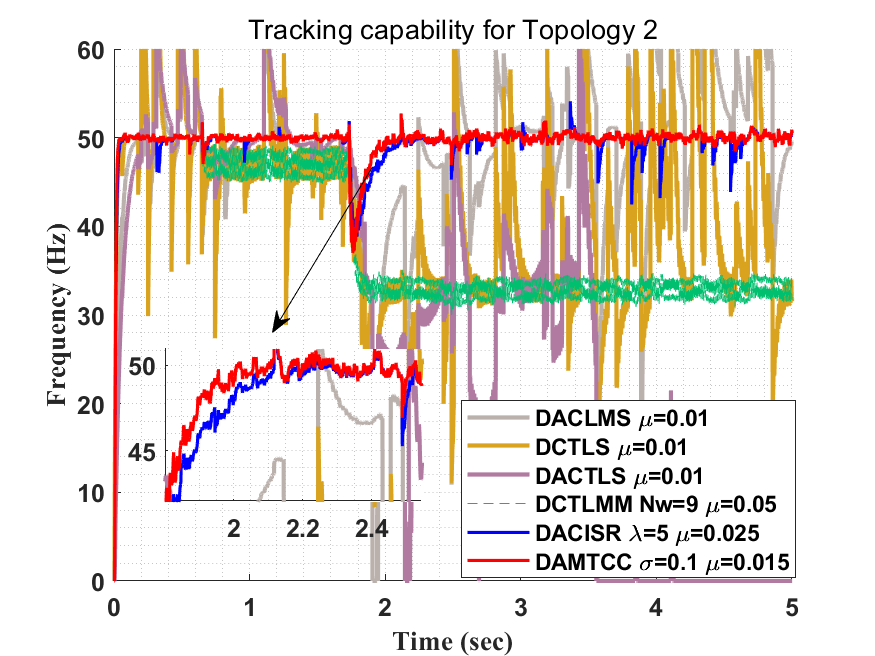}
    \end{minipage}
    }
    \caption{Perfomance comparison of tracking capability for diffusion algorithms}
\end{figure}
\vspace{-5mm}

\begin{figure}[htbp]
    \centering
    \subfigure[]{
    	\begin{minipage}[b]{.46\linewidth}
        \centering 
        
        \includegraphics[scale=0.27]{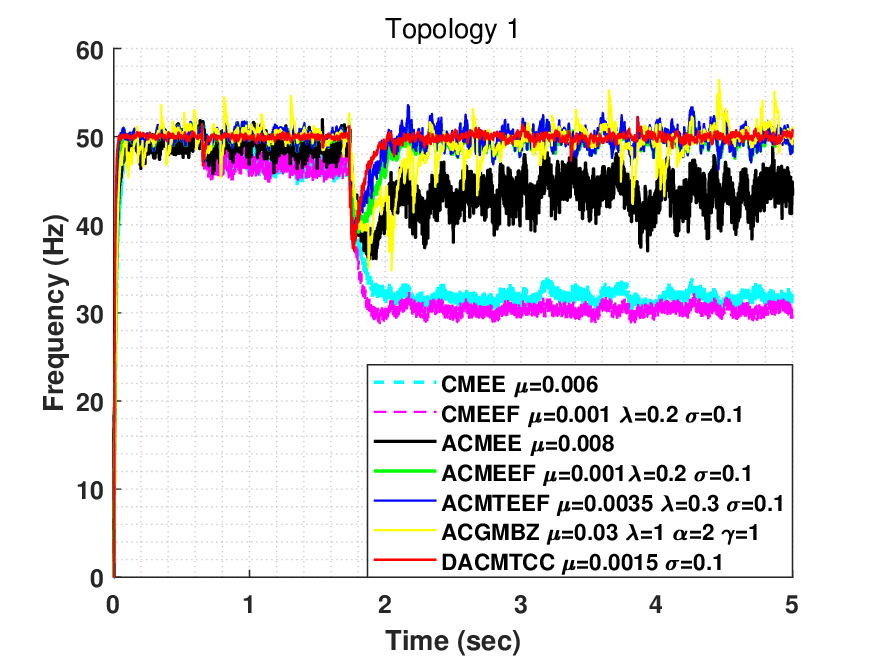} 
    \end{minipage}
    }
        \subfigure[]{
    	\begin{minipage}[b]{.47\linewidth}
        \centering
        \includegraphics[scale=0.27]{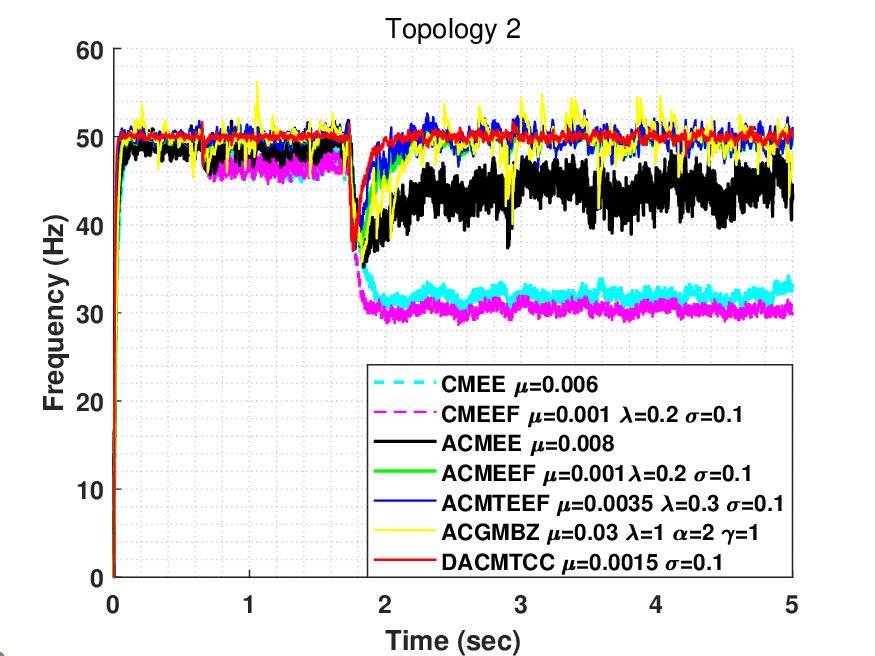}
    \end{minipage}
    }
    \caption{Perfomance comparison of tracking capability for non-diffusion algorithms}
    \vspace{-5mm}
\end{figure}

\begin{figure}[htbp]
    \centering
    \subfigure[]{
    	\begin{minipage}[b]{0.46\linewidth}
        \centering
        \includegraphics[scale=0.27]{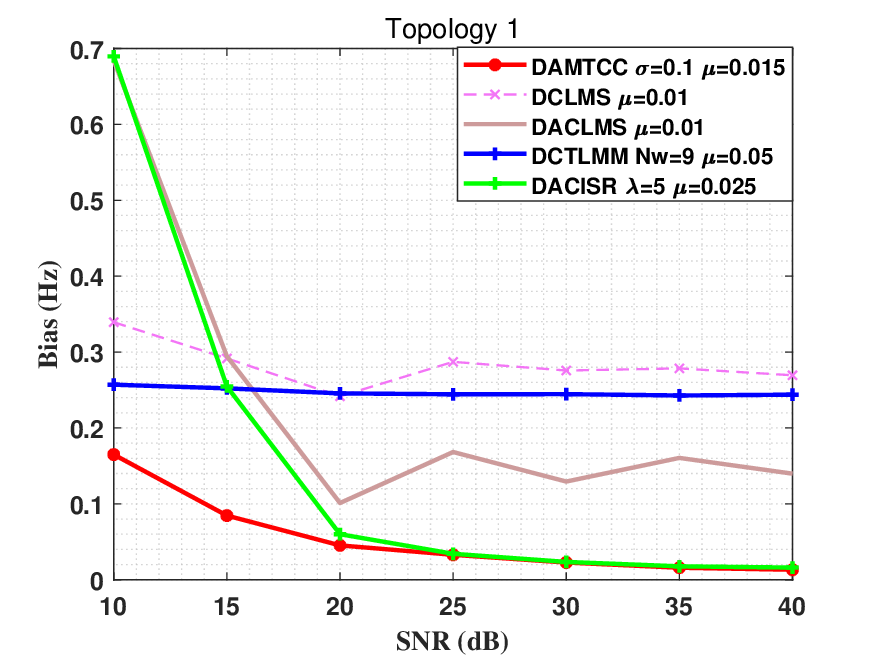} 
    \end{minipage}
    }
        \subfigure[]{
    	\begin{minipage}[b]{0.47\linewidth}
        \centering
        \includegraphics[scale=0.27]{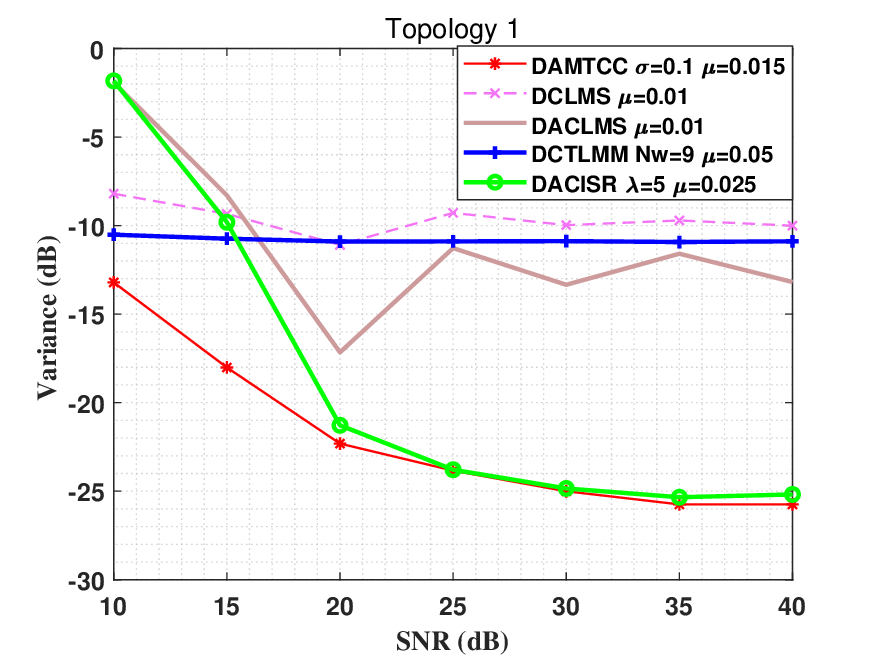}
    \end{minipage}
    }
    \caption{Steady-state  frequency estimation performance with different SNRs}
   
\end{figure}
\vspace{-3mm}

\subsection{The Algorithm Comparison for Tracking Capability }
In this section, the tracking capability of different algorithms is verified under an unbalanced three-power system.

In Fig. 3 and Fig. 4, the tracking ability of the different algorithms is depicted under the D-type  voltage sags, i.e., unbalanced three power system. The initial frequency is set to $0$Hz. As shown in Fig. 3(a) and (b), the DAMTCC algorithm has significantly better frequency tracking performance compared to the other competing algorithms.  As depicted in Fig. 4(a) and (b), to ensure fair performance comparisons, the SNR is consistently set to 40 dB for non-diffusion algorithms and all nodes in diffusion algorithms. In Fig. 4(a) and (b), leveraging the strengths of diffusion networks, the DAMTCC algorithm demonstrates superior performance across various network topologies. It achieves faster convergence than the best-performing non-diffusion algorithm, ACMTEEF, while maintaining smoother and more stable convergence curves with reduced oscillations.
\vspace{-5mm}

\subsection{Steady-state Performance with Different SNRs}
The comparison of the steady state performance for all the algorithms under topology 1 is given in Fig. 5(a) and (b).

As illustrated in Fig. 5(a), when the SNR is low, the bias value of the diffusion augmented complex inverse square root (DACISR) algorithm is large because it cannot handle the input noise. When the SNR is further increased to 25db, the curve of the DACISR algorithm coincides with the DAMTCC algorithm because the input noise is very small.  As depicted in Fig. 5(b), the algorithm proposed in this paper have better steady state performance at different SNRs. When the SNR is lower, the performance advantage is greater compared with the DACISR algorithm, and when the SNR is larger, the performance advantage is smaller. In addition, the other competing algorithms perform poorly due to 1) the influence of impulsive interference and 2) based on SL model.

\section{Conclusion}
In this letter,  the DAMTCC algorithm is proposed and applied in frequency estimation. This algorithm is able to utilize the advantages of diffusion algorithm and still maintain superior performance when the input signal contains noise and the output signal is disturbed by impulsive interference. In addition, the DAMTCC algorithm is still able to achieve adaptive frequency estimation when the power system is unbalanced. Finally, the theoretical performance is analyzed and computer simulations verify the superior performance of the algorithm.

\newpage

\small
\bibliographystyle{IEEEtran}
\bibliography{IEEEabrv,ref}

\begin{thebibliography}{10}
\providecommand{\url}[1]{#1}
\csname url@samestyle\endcsname
\providecommand{\newblock}{\relax}
\providecommand{\bibinfo}[2]{#2}
\providecommand{\BIBentrySTDinterwordspacing}{\spaceskip=0pt\relax}
\providecommand{\BIBentryALTinterwordstretchfactor}{4}
\providecommand{\BIBentryALTinterwordspacing}{\spaceskip=\fontdimen2\font plus
\BIBentryALTinterwordstretchfactor\fontdimen3\font minus \fontdimen4\font\relax}
\providecommand{\BIBforeignlanguage}[2]{{%
\expandafter\ifx\csname l@#1\endcsname\relax
\typeout{** WARNING: IEEEtran.bst: No hyphenation pattern has been}%
\typeout{** loaded for the language `#1'. Using the pattern for}%
\typeout{** the default language instead.}%
\else
\language=\csname l@#1\endcsname
\fi
#2}}
\providecommand{\BIBdecl}{\relax}
\BIBdecl

\bibitem{LMS}
S.~Haykin, \emph{Adaptive filter theory}, 2002, vol.~2.

\bibitem{CLMS}
{Y. Xia and D. P. Mandic}, ``A full mean square analysis of clms for second-order noncircular inputs,'' \emph{IEEE Trans. Signal Process}, vol.~65, no.~21, pp. 5578--5590, 2017.

\bibitem{ACLMS}
Y.~Xia and D.~P. Mandic, ``Widely linear adaptive frequency estimation of unbalanced three-phase power systems,'' \emph{IEEE Trans. Instrum. Meas}, vol.~61, no.~1, pp. 74--83, 2011.

\bibitem{ACGMBZ}
Y.~Xiao, W.~Yan, H.~Ni, B.~Chen, and W.~Wang, ``Augmented complex generalized modified blake-zisserman algorithm for adaptive frequency estimation of power system,'' \emph{IEEE Trans. Circuits Syst. II, Exp. Briefs}, 2023.

\bibitem{ACMCC}
A.~Khalili, A.~Rastegarnia, and S.~Sanei, ``Robust frequency estimation in three-phase power systems using correntropy-based adaptive filter,'' \emph{IET Sci. Meas. Technol.}, vol.~9, no.~8, pp. 928--935, 2015.

\bibitem{ACMEE}
H.~Zhao, G.~L. Nefabas, and Z.~Wang, ``Augmented complex minimum error entropy for adaptive frequency estimation of power system,'' \emph{IEEE Trans. Circuits Syst. II, Exp. Briefs}, vol.~69, no.~3, pp. 1972--1976, 2021.

\bibitem{ACMEEF}
H.~Zhao, Y.~Liu, W.~Luo, and C.~Wang, ``Augmented complex minimization of error entropy with fiducial points for power system frequency estimation,'' \emph{IEEE Trans. Circuits Syst. II, Exp. Briefs}, vol.~70, no.~6, pp. 2296--2300, 2023.

\bibitem{zhu2020robust}
Y.~Zhu, H.~Zhao, X.~Zeng, and B.~Chen, ``Robust generalized maximum correntropy criterion algorithms for active noise control,'' \emph{IEEE/ACM Transactions on Audio, Speech, and Language Processing}, vol.~28, pp. 1282--1292, 2020.

\bibitem{zhu2021cascaded}
Y.~Zhu, H.~Zhao, X.~He, Z.~Shu, and B.~Chen, ``Cascaded random fourier filter for robust nonlinear active noise control,'' \emph{IEEE/ACM Transactions on Audio, Speech, and Language Processing}, vol.~30, pp. 2188--2200, 2021.

\bibitem{EIVmodel}
T.~S{\"o}derstr{\"o}m, ``Errors-in-variables methods in system identification,'' \emph{Automatica}, vol.~43, no.~6, pp. 939--958, 2007.

\bibitem{GDTLS}
D.~Zhang, T.~Jiang, Z.~Guo, and V.~Vasil’ev, ``Real and complex solutions of the total least squares problem in commutative quaternionic theory,'' \emph{Comput. Appl. Math.}, vol.~43, no.~4, p. 235, 2024.

\bibitem{ACTLS}
Q.~Zhang, Z.~Li, H.~Jin, and X.~Chen, ``An augmented complex-valued gradient-descent total least-squares algorithm for noncircular signals,'' \emph{Signal Processing}, vol. 228, p. 109740, 2025.

\bibitem{ACMTEEF}
H.~Zhao, Z.~Cao, W.~Xu, Y.~Liu, and J.~Chen, ``Complex-valued minimum total error entropy adaptive algorithm with fiducial point for power system frequency estimation,'' \emph{IEEE Transactions on Circuits and Systems II: Express Briefs}, 2024.

\bibitem{DACLMS}
S.~Kanna, S.~P. Talebi, and D.~P. Mandic, ``Diffusion widely linear adaptive estimation of system frequency in distributed power grids,'' in \emph{Proc. IEEE Int. Energy Conf.}\hskip 1em plus 0.5em minus 0.4em\relax Cavtat, Croatia, 2014, pp. 772--778.

\bibitem{DACISR}
P.~Song, J.~Ye, K.~Yan, and Z.~Luo, ``Diffusion augmented complex inverse square root for adaptive frequency estimation over distributed networks,'' \emph{Symmetry}, vol.~16, no.~10, p. 1375, 2024.

\bibitem{17}
E.~Clarke, \emph{Circuit analysis of AC power systems: symmetrical and related components}.\hskip 1em plus 0.5em minus 0.4em\relax New York: Wiley, 1943, vol.~1.

\bibitem{18}
M.~Akke, ``Frequency estimation by demodulation of two complex signals,'' \emph{Trans. Power Del.}, vol.~12, no.~1, pp. 157--163, 1997.

\bibitem{16}
H.~Zhao, Y.~Chen, and S.~Lv, ``Robust diffusion total least mean m-estimate adaptive filtering algorithm and its performance analysis,'' \emph{IEEE Trans. Circuits Syst. II, Exp. Briefs}, vol.~69, no.~2, pp. 654--658, 2021.

\bibitem{MTCC}
G.~Qian, S.~Wang, and H.~H. Iu, ``Maximum total complex correntropy for adaptive filter,'' \emph{IEEE Transactions on Signal Processing}, vol.~68, pp. 978--989, 2020.

\bibitem{DCTLMM}
H.~Zhao, Z.~Cao, and Y.~Chen, ``Diffusion complex total least mean m-estimate adaptive algorithm for distributed networks and impulsive noise,'' \emph{IEEE Transactions on Circuits and Systems II: Express Briefs}, 2024.

\bibitem{19}
{\v{Z}}.~Ze{\v{c}}evi{\'c}, B.~Krstaji{\'c}, and T.~Popovi{\'c}, ``Improved frequency estimation in unbalanced three-phase power system using coupled orthogonal constant modulus algorithm,'' \emph{Trans. Power Del.}, vol.~32, no.~4, pp. 1809--1816, 2016.

\end{thebibliography}

\end{document}